\documentclass[sigconf,nonacm]{acmart}

\usepackage{xifthen}
\usepackage{ifthen}
\usepackage{subcaption}
\usepackage{url}
\usepackage{enumerate}
\usepackage{enumitem} 
\usepackage{algorithm}
\usepackage{algpseudocode} 
\usepackage{amsmath, amsfonts, mathtools, amsthm}
\usepackage{booktabs}
\usepackage{multirow}
\usepackage{tcolorbox}

\AtBeginDocument{%
  }

\begin{document}

\title[Assessing Attack Surfaces in Generative Search Engines]{Assessing Attack Surfaces in Generative Search Engines through Publisher Attributes: A Case Study in Political Domains}

\author{Riku Mochizuki}
\email{riku-m@keio.jp}
\affiliation{%
  \institution{Keio University/QueryLift Inc.}
  \city{Tokyo}
  \country{Japan}
}

\author{Shusuke Komatsu}
\email{shusuke.komatsu@querylift.co.jp}
\affiliation{%
    \institution{QueryLift Inc.}
  \city{Tokyo}
  \country{Japan}
}

\author{Souta Noguchi}
\email{souta.noguchi@querylift.co.jp}
\affiliation{%
  \institution{QueryLift Inc.}
  \city{Tokyo}
  \country{Japan}
}

\author{Kazuto Ataka}
\email{ataka@sfc.keio.ac.jp}
\affiliation{%
  \institution{Keio University}
  \city{Tokyo}
  \country{Japan}
}

\begin{abstract}
We characterize the attack surface of generative search engines (GSEs) against poisoning attacks in the political domain, from the perspectives of citation selection and personalization.
GSEs integrate web search and answer generation with user preferences and backgrounds using large language models (LLMs).
They play a crucial role in how users access information on the web.
Because anyone can publish content on the web, GSEs are vulnerable to poisoning attacks that manipulate citations to undermine reliable information delivery.
Existing studies on citation evaluation focus on how faithfully answers reflect cited content.
However, they leave unexamined the two critical aspects to capture the attack surface of GSEs against poisoning attacks: which publishers GSEs prefer to cite, and how personalization affects citation behavior.
To fill this gap, we introduce an evaluation framework that characterizes the attack surface of GSEs against poisoning attacks.
Our contributions are twofold: (1) we propose a novel metric, \emph{content-injection barrier}, which quantifies the difficulty of injecting arbitrary content onto the web with a given level of publisher authority; and (2) we reveal how personalization affects citation behavior by embedding user profiles into GSEs.
We conduct experiments on three major GSEs in the political domain of the United States and Japan.
Our results show that (a) the attack surface differs across GSE models; (b) the web search functionality of GSEs shapes the attack surface; (c) ruling parties have a broader attack surface than opposition parties; and (d) user profiles have little influence on the attack surface.
\end{abstract}




\maketitle

\section{Introduction}\label{sec:introduction}
Large language model (LLM) applications such as ChatGPT, Gemini, and Claude increasingly incorporate web search functions, providing internet users with new ways of accessing information~\cite{webgpt-arxiv21, geo-apmr+24, zhai2024llm-and-future-of-information-retrieval}. 
Information retrieval has become a primary use case for LLMs, with a survey showing that about $30\%$ of all prompts involve information seeking~\cite{how-people-use-chatgpt-nber25}. 
Such systems that perform web searches and generate textual answers based on users' needs and backgrounds are known as \emph{generative search engines} (GSEs)~\cite{geo-apmr+24}. 
Web search services are rapidly transitioning from traditional search engines to GSE-based systems~\cite{openai-chatgpt-search24, google-gemini-search24, anthropic-multiagent25}, changing the way users access and interpret web content.


The nature of GSEs, which cite web content published by a wide range of actors, makes them vulnerable to poisoning attacks that inject malicious content onto the web and induce attacker-intended answers~\cite{poisonedrag-arxiv24, poisoning-web-scale-training-datasets-is-practical-sp24}. 
When GSEs generate answers lacking factuality, the cause is not only LLM hallucinations~\cite{sirens-song-in-the-ai-ocean-coli25, survey-of-hallucination-in-natural-language-generation-acmcsurv23} but also the lack of factuality in the web content obtained through web searches~\cite{retrieval-helps-or-hurts-deeper-arxiv24, poisonedrag-arxiv24}.
Web content can be published and modified by any user with any intent~\cite{poisoning-web-scale-training-datasets-is-practical-sp24, poisonedrag-arxiv24}.
Prior studies formalize poisoning attacks that place disinformation and misinformation on the web to generate attacker-intended answers in systems that retrieve from external databases and generate answers, including GSEs, because these systems can cite such content and generate text that reflects it~\cite{poisoning-web-scale-training-datasets-is-practical-sp24, poisonedrag-arxiv24}.
Specific to GSEs, prior studies show patterns in styles of web content that GSEs prefer to cite in answers and propose a method to optimize their own content to increase citation exposure~\cite{geo-apmr+24, wu2025generativesearchengineslikeai}.
Although they are beneficial for primary information providers that seek to increase the exposure of their web content as citations, they can also facilitate poisoning attacks.
This indicates that systematically evaluating the citation safety of GSEs against poisoning attacks is indispensable.

Existing evaluation criteria aim to assess two aspects: (1) how faithfully the cited content is reflected in the answers, and (2) how much the cited content dominates the answers.
Several studies evaluate \emph{faithfulness} by measuring the semantic consistency between the cited web content and the generated answers~\cite{citeeval-acl25, ragas-eacl24, ares-naacl24, evaluating-verifiability-in-generative-search-engines-emnlp23, enabling-llms-to-generate-text-with-citations-emnlp23, ranking-generated-summaries-by-correctness-acl2019, evaluating-the-factual-consistency-of-abstractive-text-summarization-emnlp20, alignscore-acl23, feqa-acl20}.
Prior studies also evaluate the \emph{domination} score, which represents the proportion of cited text relative to the total text volume of the generated answer~\cite{geo-apmr+24, wu2025generativesearchengineslikeai}.
By combining these metrics, we can measure how much a cited source has the power to influence the answer to a given question.
However, these studies alone cannot capture the attack surface of GSEs against poisoning attacks.
\emph{Faithfulness} studies target systems called \emph{RAG systems}, which retrieve from curated, fixed, and closed external authoritative databases to generate answers.
In contrast, GSEs treat the web as an external database that is open to anyone and dynamically updated, allowing attackers to easily inject malicious content that GSEs can cite.
Although the \emph{domination} score is designed for the GSE context, it only measures the degree of influence of the cited content on the generated answer.
To reveal the potential attack surface, we need to consider not only the relation between cited content and answers, but also which publishers GSEs cite in those answers.
Because of these differences, existing evaluation criteria alone are insufficient to assess the attack surface.


Moreover, existing citation evaluation also does not consider the impact of personalization on citation behavior~\cite{kearney2025languagemodelschangefacts, exploreing-safery-in-personalization-arxiv2026}.
The LLM applications actively personalize inference to improve user experience~\cite{openai2024memory, anthropic2025memory, google2026personal}.
In political domains, a user's political ideology influences inference in LLMs without web search mode~\cite{political-bias-audit-arxiv-2026, miyazaki_hall_2026_japan_ai_voting_advice}.
However, how personalization influences the selection of citations in GSEs remains insufficiently explored.
To reveal the attack surface of GSEs under real deployment scenarios, we need to consider the effects of personalization on their citation behavior.

To address these gaps, we introduce a novel framework to characterize the potential attack surface of GSEs against poisoning attacks.
Our framework classifies each citation of answers by GSEs into \emph{content-injection barrier}, categorized as low, medium, or high, based on the practical difficulty of publishing content on the web with a given level of publisher authority.
In addition, the framework evaluates the faithfulness score and domination score across different content-injection barriers to validate the identified potential attack surfaces in specific scenarios.

We conduct experiments on three major GSEs (OpenAI GPT-5, Claude Sonnet 4, and Gemini Flash 2.5 with search mode enabled) in the United States and Japan.
We use $33$ question templates across $13$ parties, $3$ user profiles, and $3$ GSE models, totaling $3{,}861$ generated answers.
To retrieve GSE answers, we set a user profile that varies the user's level of political knowledge and ideology, and pose political questions to GSEs.
This enables systematic analysis of personalization's effect on citation behavior.

Our experiments reveal four findings.
First, each GSE has a distinct attack surface: some GSEs concentrate citations on primary information sources and exclude low-barrier sources, while others cite low-barrier sources more frequently.
Second, when citations contain few primary sources, web search results tend to include fewer primary sources and more low-barrier sources.
This suggests that the web search functionality of GSEs (e.g., query generation by LLMs) shapes the potential attack surface, rather than the LLM's preference for which sources from web search results to cite.
Third, answers to ruling party questions rely more on sources with lower content-injection barriers than answers to opposition party questions, making the ruling side more vulnerable.
Lastly, in contrast to biases arising from the LLM's internal knowledge, user profiles have little influence on the citation barrier pattern, faithfulness, or domination score across content-injection barriers.

Our main contributions are as follows:
\begin{itemize}
    \item We propose an evaluation framework of GSEs to characterize the potential attack surface against poisoning attacks.
    As a technical contribution, the framework classifies citations of answers by GSEs into \emph{content-injection barrier}, and evaluates the potential attack surface along three axes: barrier distribution, faithfulness, and domination score.
    \item We conduct experiments on three major GSEs in the United States and Japan in the political domain, using $33$ question templates across $13$ parties, $3$ user profiles, and $3$ GSE models for a total of $3{,}861$ generated answers.
    \item We find that the potential attack surface differs by GSE model and by whether the party in question is in the ruling or opposition position, but not by user profile.
\end{itemize}

\section{Generative Search Engine}
\label{sec:generative_search_engine}
This section introduces Generative Search Engine (GSE) and its system model.
GSEs are retrieval-and-generation systems that retrieve information from the web and generate synthesized answers with user intent and backgrounds using LLMs~\cite{geo-apmr+24}.

Aggarwal et al.\ formalize GSEs and provide a system model~\cite{geo-apmr+24}.
A GSE is formalized as a function \( f_{GSE}^{model} \) that takes a user question \( q_u \) and personalization information \( P_U \) as inputs and generates a textual answer \( r \):
$f_{GSE}^{model} := (q_u, P_U) \rightarrow r$.

GSEs consist of two components: content retrieval and answer generation.
Content retrieval collects the information necessary to generate answers from the web.
The user query \( q_u \) is converted by LLMs into multiple queries \( Q^{\prime} = \{q_1, q_2, \cdots, q_n\} \) for the web search.
These queries are sent to a search engine \( SE \), and GSEs obtain a set of web sources \( S = \{s_1, s_2, \cdots, s_m\} \) from the search results.
Each result set in $S$ for a web search query \( q_i \) typically consists of the top \( k \) web sources ranked by the search engine's metrics~\cite{poisonedrag-arxiv24, modern-information-retrieval-a-brief-overview-3320, geo-apmr+24}.

Answer generation cites the web sources obtained in the content retrieval phase and outputs the textual answer \( r \).
Web sources \( S \) are converted into a summary set \( Sum = \{Sum_1, Sum_2, \cdots, Sum_m\} \) that extracts and summarizes the content of web sources obtained in the content retrieval phase.
LLMs then generate the final answer \( r \) from the summary set \( Sum \) while citing web sources \( S \).
The answer \( r \) consists of \( k \) sentences \( \{l_1, l_2, \cdots, l_k\} \), and each sentence \( l_i \) is associated with a citation set \( C_i \subseteq S, C_i = \{c_1, c_2, \cdots, c_l\} \).
Although GSEs ideally have one or more citations for each sentence, there are cases where \( C_i = \emptyset \), meaning there are no citations for the corresponding sentence.

A core technology of GSEs is Retrieval-Augmented Generation (RAG), which retrieves content relevant to the user query from outside of LLMs and incorporates it into LLM inference~\cite{internet-augmented-language-models-fewshot-arxiv22, internet-augmented-dialogue-generation-arxiv21, rag-kdd24}.
Information retrieval systems that use RAG are called \emph{RAG systems}~\cite{mirage-acl24, lála2023paperqa, almanac-nejmai-2024}.

However, several studies of \emph{RAG systems} implicitly focus on systems that retrieve from authoritative, fixed, curated, and closed external databases defined by developers~\cite{mirage-acl24, lála2023paperqa, almanac-nejmai-2024}.
While GSEs similarly use the web as an external database, the nature of this database differs from that of these RAG systems.
The web targeted by GSEs is open, allowing anyone to freely publish information, including malicious content, and is dynamic, with information being updated as needed.
Therefore, unlike these RAG systems, GSEs use search engines (SE) for retrieval from the external database (the web).
Our study distinguishes GSEs and RAG systems based on the differences in external database characteristics.

\section{Related Work and Motivation}
\label{sec:related}
This section introduces poisoning attacks on \emph{RAG systems}, summarizes key evaluation criteria, and highlights the challenges we identify for GSEs.

\subsection{Vulnerabilities of GSEs}
We introduce poisoning attacks on GSEs.
Zou et al.~\cite{poisonedrag-arxiv24} reveal and formalize an attack method called \emph{PoisonedRAG} that exploits \emph{RAG systems} to generate attacker-intended answers by placing malicious content into external databases.
Even if attackers can inject a small volume of malicious text into external databases, \emph{RAG systems} can produce attacker-intended answers for specific questions.
This malicious content includes text that attackers want displayed as answers to specific questions, especially false information such as ``\textit{OpenAI's CEO is Tim Cook.}''
PoisonedRAG attacks target closed-ended questions like ``\textit{Who is the CEO of OpenAI?}'' rather than open-ended questions like ``\textit{What are the latest trends in AI?}''~\cite{poisonedrag-arxiv24}.

In fact, several studies support the effectiveness of this attack.
Chen et al.~\cite{chen2022rich} show that in \emph{RAG systems}, when content containing two different claims (e.g., \textit{OpenAI's CEO is Tim Cook.} and \textit{OpenAI's CEO is Sam Altman.}) is placed and answers are generated for each claim, the answers may include each respective claim.
Wang et al.~\cite{wang2025ragconflictevidence} and Liu et al.~\cite{liu2025conflicts} demonstrate that even when there is an imbalance in the volume of supporting documents for each claim, \emph{RAG systems} tend to favor the answer with more supporting evidence, yet the underrepresented claim can still appear in responses.
These studies support the effectiveness of PoisonedRAG attacks and demonstrate that RAG may cite malicious content even when attackers place only a small volume of it.

To succeed in poisoning attacks, the malicious content must satisfy two conditions: \emph{Retrieval Condition} and \emph{Generation Condition}.
Retrieval Condition means that malicious content (such as text or HTML placed by attackers) is selected as top-$ k $ relevant content through retrieval tasks for target questions.
In GSEs, Retrieval Condition means that content on the web is retrieved as web pages ranked highly in search engine results.
Generation Condition means that the content is used in LLM inference and reflected in the generated answers.
Aggarwal et al.~\cite{geo-apmr+24} propose \emph{generative engine optimization} (GEO), a method that optimizes website text structure and content from the perspective of Generation Condition to increase the exposure of web page content as citations in answers when the website passes the Retrieval Condition.
GEO provides theoretical evidence that GSEs are vulnerable to poisoning attacks.


\subsection{Citation Evaluation}
\label{subsec:citation_evaluation}
We first introduce seminal studies that aim to evaluate retrieval-and-generation systems.
These studies evaluate a citation's influence on the generated answer mainly from two aspects: how much it shapes the answer and how faithfully it is reflected in the answer's sentences.
To evaluate these systems, the key evaluation metrics are \emph{faithfulness} and \emph{domination}.

\emph{Faithfulness} measures how accurately the generated answer reflects the content of its citations~\cite{citeeval-acl25, ragas-eacl24, ares-naacl24, evaluating-verifiability-in-generative-search-engines-emnlp23, enabling-llms-to-generate-text-with-citations-emnlp23, ranking-generated-summaries-by-correctness-acl2019, evaluating-the-factual-consistency-of-abstractive-text-summarization-emnlp20, alignscore-acl23, feqa-acl20}.
Faithfulness is typically computed by three methods: entailment-based~\cite{ranking-generated-summaries-by-correctness-acl2019, true-reevaluating-factual-consistency-arxiv22, evaluating-the-factual-consistency-of-abstractive-text-summarization-emnlp20, alignscore-acl23}, similarity-based~\cite{bertscore-arxiv20, bartscore-neurips21}, and QA-based~\cite{feqa-acl20}.
Similarity-based methods are most appropriate for calculating reflection accuracy.
Similarity-based methods quantitatively measure semantic similarity between two texts and rely on neural encoder models.
Zhang et al.\ show that similarity-based methods achieve higher precision than entailment-based and QA-based methods in tasks evaluating consistency between long documents and their summaries~\cite{finegrainedcitationevaluationgenerated-arxiv24}.
This approach aligns with GSE behavior of summarizing web page content retrieved from search results and generating answers with citations.
Citation evaluation applies these faithfulness evaluation methods to each sentence $ l_i $ of generated answer $ r $ with citations $ C_i $, and evaluates reflection accuracy for the entire answer $ r $.

\emph{Domination} measures the proportion of cited text relative to the total text volume of the generated answer~\cite{geo-apmr+24, wu2025generativesearchengineslikeai}.
For each citation $ C_i $, they count the number of words in the sentences to which $ C_i $ is attached and report this count as a proportion of the total number of words in the generated answer $ r $.
This is a major metric for evaluating the influence of citations in GSEs.

However, they do not assess the authority and trustworthiness of cited content, as they disregard the publisher of those sources.
Such content-driven evaluations limit their applicability to GSEs for revealing the potential attack surface of poisoning attacks.
In \emph{RAG systems}, poisoning attack studies focus on how much malicious content attackers need to inject into external databases to succeed in a poisoning attack.
In contrast, GSEs retrieve information from the open web, where attackers can inject malicious content more easily than in \emph{RAG systems}.
This ease of injection can enable a successful poisoning attack as previously described.
The main concern shifts from preventing malicious content from existing in external databases to preventing the system from incorporating even a small amount of malicious content injected by attackers during web searches.
Unlike \emph{RAG systems}, GSEs operate under the assumption that malicious content is always present in their external databases (the web), and therefore existing evaluation methods alone are insufficient to assess the safety and reliability of GSEs.

\subsection{Authority and Trustworthiness}
Several studies propose methods to improve the inclusion of authoritative sources in citations generated by \emph{RAG systems} and GSEs~\cite{zhou2025trustragenhancingrobustnesstrustworthiness, lee2026relevanceauthorityauthorityawaregenerative}.
TrustRAG~\cite{zhou2025trustragenhancingrobustnesstrustworthiness} identifies and removes malicious documents during content retrieval by clustering candidate documents and matching them with the internal knowledge of the LLM.
AuthGR~\cite{lee2026relevanceauthorityauthorityawaregenerative} computes an authority score for web pages using a vision-language model that evaluates both the textual content of the page and visual features extracted from page screenshots, such as layout quality, design, and advertisement density. 
The system prefers to cite web pages with high authority scores.

However, evaluating only the citation content itself is insufficient for assessing the safety of GSEs under poisoning attacks.
As described in Section~\ref{subsec:citation_evaluation}, because anyone can publish content on the web, attackers can deliberately optimize the scores of content-based evaluation metrics while still injecting malicious information into the web.
In fact, the ``Doppelganger'' operation reported by EU DisinfoLab~\cite{eudisinfolab2022doppelganger,foreignaffairs2024doppelganger} shows that more than $700$ fake news websites were created by imitating designs of legitimate media, and their content was distributed while disguised as legitimate media.
Such legitimate-looking malicious content undermines approaches that evaluate only the content itself, causing both false negatives and false positives.
Therefore, for GSEs, we need to evaluate not only the citation content itself, but also the authority and trustworthiness of content publishers.

Furthermore, these methods typically assume the presence of poisoning attacks and primarily focus on developing defenses or system improvements. 
However, they do not provide a systematic understanding or measurement of the actual extent of citation selection bias and vulnerability found in production GSEs.

\subsection{Personalization of LLMs}
\label{sec:related_works_personalization}
Personalization is a key factor in improving the user experience of chat-based applications~\cite{openai2024memory, anthropic2025memory, google2026personal} and is a recognized research direction for language models~\cite{salemi2024lamp, mysore2024pearl}.
In personalization, the general approach is to embed the user's attribute information together with the user's question.
This approach enables the applications to generate answers that more accurately reflect the user's preferences, background, and needs.

Several studies investigate how personalization affects inferences based on the internal knowledge of LLMs.
Kearney et al.~\cite{kearney2025languagemodelschangefacts} show that across several domains, answers change systematically depending on user attributes such as race, gender, and age.
The influence of user attributes also extends to tasks that have objectively correct answers.
Vijjini et al.~\cite{exploreing-safery-in-personalization-arxiv2026} show that LLM response performance varies systematically even in tasks such as mathematical reasoning, general knowledge, and programming.

This effect also extends to the user's political ideology.
Tornberg et al.~\cite{political-bias-audit-arxiv-2026} analyze political ideology bias in representative and widely used LLMs.
The study shows that LLMs produce opposition-leaning answers by default and LLMs shift strongly toward the ruling party for conservative users but only slightly toward the opposition party for progressive users, showing an asymmetric response to personalization.
Miyazaki et al.~\cite{miyazaki_hall_2026_japan_ai_voting_advice} show that a similar pattern holds in the Japanese language and Japanese political context.
They show that when users do not indicate their policy positions, LLM answers lean toward the ruling party, whereas when users explicitly indicate their political stance, the answers shift substantially in that direction.

However, these prior studies focus on biases that arise from the internal knowledge of LLMs and do not examine biases in GSEs. 
To the best of our knowledge, no prior study has examined how personalization affects the citation behavior of GSEs, including aspects such as source authority and trustworthiness, and domination and faithfulness of citations.

\subsection{Research Motivation}
To understand the potential attack surface of GSEs, our goal is to build a novel evaluation framework. 
Our framework introduces new evaluation criteria centered on the publisher attributes of citations in generated answers, enabling a systematic assessment of the underlying attack surfaces and the resilience of GSEs against poisoning attacks.
Note that our framework does not aim to identify actual malicious content, nor to demonstrate practical attacks that could exploit such surfaces; its sole purpose is to evaluate potential attack surfaces.

Our framework addresses the following research questions:
\begin{description}
  \item[\textbf{\textit{R-1}}] Which publishers' content do GSEs prefer to cite, and does this reveal any bias in citation selection from the perspective of publishers?
  \item[\textbf{\textit{R-2}}] Which publishers' content has the power to influence the domination and faithfulness of citations?
  \item[\textbf{\textit{R-3}}] How does personalization affect the citation behavior of GSEs?
\end{description}

We apply our framework to the political domain, where GSEs increasingly mediate access to information essential to democratic processes.
Because the flow of information from political parties to citizens is a cornerstone of democratic electoral processes~\cite{the-trust-gap-young-peoples-tactics-for-assessing-the-reliability-of-political-news-ijpp22, kovach2007elements}, GSEs can substantially influence citizens' decision-making in elections~\cite{fisher2026biasedaiinfluencepolitical}.
Especially, focusing on the task of voters gathering information about specific political parties, we characterize the potential attack surface across political parties, ideologies, countries, and voter profiles (e.g., prior political knowledge and ideology).
Based on this analysis, we discuss where attackers could distort political information delivered through GSEs.

\section{Proposed Framework}

\begin{figure*}[t]
  \centering
  \includegraphics[width=\textwidth]{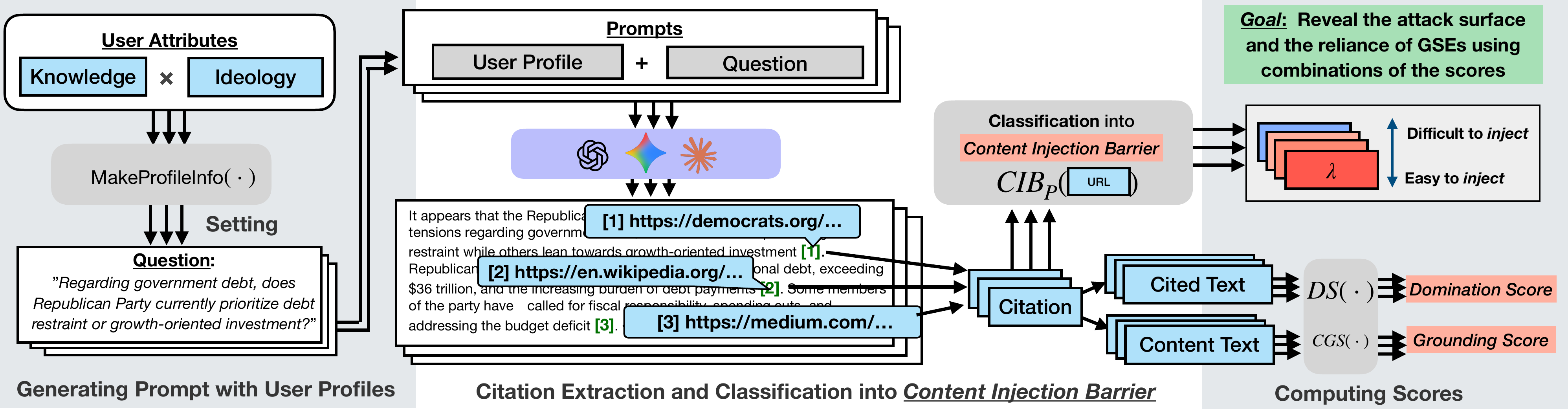}
  \caption{Overview of our framework. Given a user profile and question sets, the GSE generates an answer with citations, which are classified into content-injection barriers. 
  We then identify the attack surface along three axes: (1) barrier distribution, (2) domination score across barriers, and (3) grounding score across barriers.}
  \label{fig:overview}
\end{figure*}

In this section, we present a framework for revealing the potential attack surface of GSEs.
As a technical contribution, we introduce \emph{content-injection barriers} of each publisher in citations to characterize the practical difficulty of publishing content on the web with specific publisher attributes and authority.
In this framework, we characterize potential attack surfaces along three axes: (1) barrier distribution, (2) domination score across barriers, and (3) grounding score across barriers.
Figure~\ref{fig:overview} illustrates the overall pipeline.

Our framework is designed not to directly measure the success rate of poisoning attacks, but rather to identify citation behaviors that constitute potential attack surfaces.
Estimating the attack surface allows us to characterize one of the vulnerabilities of GSEs to poisoning attacks.

\subsection{Setting User Profiles into GSEs}
As one form of personalization, we set a user profile, which assembles a user's attributes, to GSEs.
We select a user attribute set $\mathcal{A}$ that influences the retrieval and generation tasks of GSEs.
We then build a set of user profiles from their combinations $\mathcal{U} = \textsc{Combinations}(\mathcal{A})$.

To provide a user profile $u \in \mathcal{U}$ to GSEs, we introduce a storytelling function $\textsc{MakeProfileInfo}(u)$ that converts $u$ into personalization information $P_U$ to be set in the GSE.
Note that $P_U$ is used as part of the system prompt, and $f_{GSE}^{model}: (q_u, P_U) \rightarrow r$ from Section~\ref{sec:generative_search_engine} models the personalized information retrieval and answer generation.
Following Lutz et al.~\cite{lutz2025promptmakespersonasystematic} and related prior studies~\cite{argyle2023outofone, santurkar2023whose}, we combine interview formats and name-based priming to accurately set user profiles in GSEs.

\subsection{Injection Barrier}
We introduce the \emph{content-injection barrier}, which refers to the practical difficulty of publishing content on the web based on specific publisher attributes and authority. 
We define a classifier $\textsc{CIB}_{\mathcal{P}}(c)$ for categorizing citations in answers generated by GSEs into content-injection barrier categories: \emph{low}, \emph{medium}, \emph{high}, and \emph{primary}, where $\mathcal{P}$ denotes the target political party and $c$ denotes the citation.
These barriers indicate the level of publisher authority and editorial process that an attacker must obtain to successfully inject malicious content to influence GSE answers.

We perform multi-stage classification to ensure interpretability of the classification results.
Because content-injection barriers are abstract high-level categories, we ensure the validity and interpretability of this categorization by first classifying publishers into primary and secondary information sources, then categorizing secondary sources into fine-grained publisher categories, and finally deriving content-injection barriers from these categories.

\textbf{Primary Information Identification.}
We identify primary information sources, which are domains officially operated by the target party $\mathcal{P}$.
First, we define the target domain set $\mathcal{D}_{\mathcal{P}}$ that enumerates the official domains of $\mathcal{P}$ (e.g., when $\mathcal{P}$ is the U.S. Democratic Party, $\mathcal{D}_{\mathcal{P}} = \{\text{democrats.org}, \dots\}$).
Let $d_c = \textsc{Domain}(c)$ be a function that extracts the domain from each citation $c \in C$ in a generated answer $r$ of a GSE.
We check whether $\mathcal{D}_{\mathcal{P}}$ includes $d_c$.
When $d_c$ is included in $\mathcal{D}_{\mathcal{P}}$, we classify the citation as a primary information source.
Here, $\textsc{IsPrimary}_{\mathcal{P}}(c)$ outputs $1$ if $d_c$ is included in $\mathcal{D}_{\mathcal{P}}$ and $\bot$ otherwise.

\textbf{Secondary Information Category Classification.}
We categorize secondary information into fine-grained publisher categories, which serve as intermediate categories for calculating content-injection barriers.
We prepare a publisher category set \\$\Pi = \{\pi_1, \pi_2, \cdots, \pi_p\}$.
Here, $\textsc{Cat}(c)$ maps citation $c$ to a category $\pi_i \in \Pi$ ($i \in \{1, 2, \cdots, p\}$).
In our experiments in Section~\ref{sec:experiment}, we use labels such as ``Party'', ``Media'', ``Platform'', ``Owned'', ``Academia'', ``Non-media-industry'', and ``Government.''

To build the function $\textsc{Cat}(c)$, we adopt a hybrid strategy combining automatic category classification using LLM-as-a-Judge~\cite{survey-llm-as-a-judge-arxiv24,judging-llm-as-a-judge-with-mt-bench-and-chatbot-arena-arxiv23} and manual category classification.
We prepare two GSE models (ideally two models from different providers such as GPT and Gemini with web search mode enabled) and use them to identify publishers from domains and WHOIS information.
Combining multiple GSE models can reduce potential biases of single GSE models and increase classifier accuracy~\cite{judging-llm-as-a-judge-with-mt-bench-and-chatbot-arena-arxiv23}.
When classification results from the two GSE models agree, we adopt that result directly; when they disagree, we perform manual classification.
In our experiments, the authors make the final determination based on company information from the website and domain registration data.
We employ zero-shot classification~\cite{radford2019language} following Kostina et al.~\cite{kostina2025largelanguagemodelstext}, who demonstrate high performance without training examples.
For each citation $c \in C$, the prompt receives two inputs: $\textsc{Url}(c)$ extracts the complete URL indicating the publisher's domain and path, and $\textsc{Whois}(\textsc{Url}(c))$ retrieves domain registration data including ownership and organizational information.

\textbf{Content-Injection Barrier Classification.}
Finally, we classify citations into content-injection barriers based on the primary information identification and secondary information category classification results.
We derive content-injection barriers from these categories.
Here, $\textsc{CatBar}_{\mathcal{P}}(\pi)$ maps publisher category $\pi \in \Pi$ to a content-injection barrier $\lambda_j \in \Lambda = \{\lambda_1, \lambda_2, \cdots, \lambda_q\}$.
The overall classifier is then defined as the composition $\textsc{CIB}_{\mathcal{P}}(c) = \lambda_{\text{prim}}$ if $\textsc{IsPrimary}_{\mathcal{P}}(c) = 1$, and $\textsc{CIB}_{\mathcal{P}}(c) = \textsc{CatBar}_{\mathcal{P}}(\textsc{Cat}(c))$ otherwise.
We define five content-injection barriers $\Lambda$: \emph{primary sources}, \emph{opponent sources}, \emph{low-barrier sources}, \emph{medium-barrier sources}, and \emph{high-barrier sources}.
\emph{Primary sources} are domains owned by the party referenced in the question. If $\textsc{IsPrimary}_{\mathcal{P}}(c) = 1$, then $c$ is classified as a primary information source.
\emph{Opponent sources} are domains operated by rival parties. If $\textsc{IsPrimary}_{\mathcal{P}}(c) = \bot$ and $\textsc{Cat}(c)$ is ``Party'' in our experiment, then $c$ is classified as an opponent information source.
\emph{Low-barrier sources} are domains where users or owners can freely publish or edit content (``Platform'' and ``Owned'' in our experiments).
The political influence operations described in Section~\ref{sec:related}, including troll accounts on social media platforms~\cite{zannettou2018disinformation, cinus2025cross, nippon2025russian}, fake news websites mimicking legitimate media~\cite{eudisinfolab2022doppelganger, foreignaffairs2024doppelganger}, and LLM-generated false claims~\cite{openai2026disrupting}, are all published through low-barrier sources, as these platforms and self-owned domains require no editorial oversight.
\emph{Medium-barrier sources} are domains owned by organizations or companies with editorial processes where journalism bias or interests may appear (``Media'' and ``Non-media-industry'' in our experiments).
\emph{High-barrier sources} are domains where authors are required to remain neutral and objective and manipulation is difficult (``Academia'' and ``Government'' in our experiments).

An increasing proportion of lower-barrier citations suggests that it is easier for attackers to successfully influence generated answers by injecting malicious content on the web.
Therefore, the degree of content-injection barriers serves as an indicator inversely correlated with the difficulty of successful attacks.

\subsection{Domination Score}
Our framework computes the \emph{Domination Score} ($\textsc{DS}$) for each generated answer.
It quantifies the proportion of cited text relative to the total text volume of the generated answer.
This metric has been one of the primary evaluation criteria in the context of GEO, originally defined as a word-count-based measure~\cite{geo-apmr+24}.
However, decomposing text into words is language-dependent and biased toward English; thus, we redesign it on a character-count basis and refer to it as the \emph{Domination Score}.
The domination score is computed as $\textsc{DS}(r) = \sum_{c \in C} \textsc{CharCount}(c) / \textsc{CharCount}(r)$, where $\textsc{CharCount}(\cdot)$ returns the number of characters in the given text.
By aggregating the domination score for each content-injection barrier, we analyze the impact of each barrier's content on the generated answers.

\subsection{Citation Grounding Score}
\label{subsec:citation_grounding}
Our framework computes the \emph{Citation Grounding Score} ($\textsc{CGS}$) for each citation.
It quantifies how strongly the answer sentences are reflected in the cited content, which relates to faithfulness evaluation.
We introduce a scoring function $\textsc{CGS}(c, r)$ that measures the citation grounding score for each citation $c$ and answer $r$.
By aggregating the citation grounding score for each content-injection barrier, we analyze the effectiveness of each barrier's content.

We decompose the computation into three steps:

\textbf{Decomposing Answers and Citations into Sets of Sentences.}
First, we decompose answer $r$ into a sentence set $L = \{l_1, l_2, \ldots, l_k\}$ and each citation $c_i \in C$ into a sentence set $L_{c_i} = \{l_{i,1}, l_{i,2}, \ldots, l_{i,n_i}\}$.
Because some sentences of answer $r$ have citation labels attached and the semantics of each answer sentence is independent, we decompose answer $r$ into sentences to evaluate grounding.

\textbf{Sentence-level Similarity.}
We measure semantic similarity between each answer sentence in $L = \{l_1, l_2, \ldots, l_k\}$ and each citation sentence in $L_{c_i} = \{l_{i,1}, l_{i,2}, \ldots, l_{i,n_i}\}$ for each citation $c_i \in C$.
To measure similarity between two sentences, we introduce the similarity function $\mathrm{sim}(x, y)$ between two sentences $x$ and $y$.
The function $\mathrm{sim}(x, y)$ returns a similarity score between the two sentences $x$ and $y$ as a value in the range of $-1$ to $1$.
Here, a similarity value closer to $-1$ indicates lower semantic similarity, whereas a value closer to $1$ indicates higher semantic similarity.
To calculate the similarity, we employ Sentence-BERT~\cite{reimers2019sbert}, which converts each sentence into dense embedding vectors and computes their semantic similarity through cosine similarity.
In our experiments in Section~\ref{sec:experiment}, we use the pre-trained Sentence-BERT model ``stsb-xlm-r-multilingual''~\cite{reimers2019sbert}, which has been trained on multilingual semantic textual similarity tasks.

\textbf{Maximum-grounding Aggregation.}
GSEs often use only a few sentences within long citation texts to generate a particular answer sentence; thus, we aggregate the sentence-level similarities by taking the maximum between the answer sentence set $L$ and citation $c_i$ to reveal how strongly a citation grounds the answer.
Taking the maximum identifies the most relevant part of the citation text to the answer content.
Concretely, we define $\textsc{MaxSim}(L, c_i) = \max_{j \in \{1,\ldots,k\},\, m \in \{1,\ldots,n_i\}} \mathrm{sim}(l_j, l_{i,m})$, where $l_j \in L$ and $l_{i,m} \in L_{c_i}$.
In computing this maximum, we evaluate $\mathrm{sim}(\cdot, \cdot)$ for every answer-citation sentence pair $(l_j, l_{i,m})$ within citation $c_i$ and adopt the highest similarity value across all combinations.


\section{Experiment}
\label{sec:experiment}
We apply our framework to questions about U.S. and Japanese political parties on major GSE models.

\subsection{Experimental Setup}
We provide details of our experimental setup, including the target GSE models, user profiles, questions, and political parties.
To ensure the reproducibility of our study, the question-answer datasets and the prompts used for classification are available in our repository~\footnote{Our Repository: \url{https://github.com/mzhkz/QL_Research_CIKM26_Artifacts}}.

\textbf{Target GSE Models.}
We employ three GSE models for answer generation: OpenAI GPT-5, Claude Sonnet $4$ (claude-sonnet-4-20250514), and Gemini Flash $2.5$ (gemini-2.5-flash) with web search mode enabled.
According to surveys~\cite{a16z2025stateconsumerai}, closed models dominate the LLM market in both enterprise and consumer segments, with ChatGPT, Claude, and Gemini together accounting for $88\%$ of spending share, while open-source models account for only $11\%$.
These models are developed in the U.S., but they dominate usage in Japan as well according to GMO Research \& AI~\cite{gmo-research2025japanai}.
Furthermore, Google AI Overview, which integrates Google Search by default, relies on the closed model Gemini~\cite{stein2025expandingaioverviewsaimode}.
In our experiments, we set the temperature of each GSE model to $1.0$, and the answers were obtained in May $2026$.

\textbf{User Profiles.}
We design three user profiles based on two user attributes: political knowledge, which represents the user's familiarity with political issues and takes the levels \emph{Ignorant} and \emph{High}; and political ideology, which represents the user's political leanings and takes the levels \emph{Progressive}, \emph{Conservative}, and \emph{Neutral}.

Using these attributes, we construct three user profile combinations: \emph{Ignorant-Neutral}, \emph{High-Progressive}, and \emph{High-Conservative}.
The \emph{Ignorant-Neutral} profile represents a user with no political knowledge and a neutral ideology; the \emph{High-Progressive} profile represents a user with high political knowledge and a progressive ideology; and the \emph{High-Conservative} profile represents a user with high political knowledge and a conservative ideology.
The user profiles \emph{Ignorant-Progressive} and \emph{Ignorant-Conservative} are not introduced, because the political ideology attribute is not applicable to users without political knowledge; for the same reason, \emph{High-Neutral} is not introduced.

\textbf{Questions.}
We design political questions on topics common to each country to elicit policy positions in a politically neutral manner.
Our question design focuses on whether GSEs cite content articulating positions on the target topics, rather than on questions tailored to a specific party.
All questions are closed-ended and include a party name following PoisonedRAG (e.g., ``\emph{What is \{PARTY\}'s current position on economic regulation and free markets?}'').

To avoid bias, we build the question set on MARPOR (Manifesto Project)~\cite{lehmann2025manifesto}, an established political-science framework for classifying party manifestos.
MARPOR has a hierarchical structure consisting of seven \emph{policy domains}, each of which is composed of multiple \emph{topics}.
We select four voter-facing MARPOR domains most relevant to everyday political judgment: \emph{External Relations}, \emph{Economy}, \emph{Welfare and Quality of Life}, and \emph{Fabric of Society}.
We then select $11$ topics on which all target parties take an explicit stance.
Using these $11$ topics as input, we generate questions using three closed-weight LLMs and finally obtain $33$ question templates for asking the position of each party in each country.
Questions and answers are written in English for U.S. parties and in Japanese for Japanese parties.
Each question is instantiated per party, yielding $33 \times 5 = 165$ queries for the U.S. and $33 \times 8 = 264$ for Japan.
Combined with three user profiles and three GSE models, this produces $165 \times 3 \times 3 = 1{,}485$ U.S. queries and $264 \times 3 \times 3 = 2{,}376$ Japan queries, for a total of $3{,}861$ generated answers.

Following the DETAIL framework~\cite{kim2025mattersmeasuringimpactprompt}, we fix question specificity at the least specific domain-level (asking about a stance on an entire policy domain without proper nouns or concrete figures).
It prevents questions and citations from being skewed toward topics that specific publishers actively promote, and reflects the abstraction level typical of voters' everyday political judgments.

\textbf{Target Political Parties.}
We target political parties satisfying each country's requirements for national political parties as primary sources.
Specifically, in Japan we target eight parties.
The ruling parties are ``Liberal Democratic Party'' and ``Japan Innovation Party'', and the opposition parties are ``Centrist Reform Coalition'', ``Democratic Party for the People'', ``Japanese Communist Party'', ``Reiwa Shinsengumi'', ``Sanseito'', and ``Conservative Party of Japan''.
In the U.S., we target five parties. The ruling party is ``Republican Party'', and the opposition parties are ``Democratic Party'', ``Green Party'', ``Libertarian Party'', and ``Constitution Party.''

\subsection{Experimental Results}

\begin{figure}[t]
  \centering
  \includegraphics[width=\columnwidth]{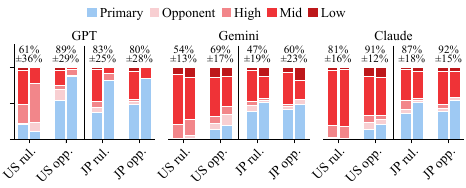}
  \caption{Barrier distribution by party group. Each cell has two stacked bars: the left shows the barrier distribution of web search results, and the right shows that of GSE citations.}
  \label{fig:citation-by-group}
\end{figure}

\begin{figure*}[t]
  \centering
  \includegraphics[width=\textwidth]{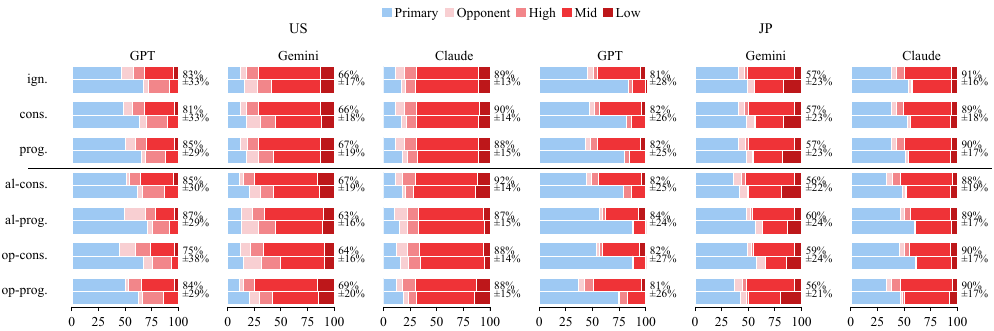}
  \caption{Barrier distribution by user profile. Each bar shows the barrier distribution of GSE citations for each user profile.}
  \label{fig:citation-by-persona}
\end{figure*}

We analyze the citation behavior of GSEs along three axes: (1) the content-injection barrier distribution of citations, (2) the domination score distribution across barriers, and (3) the citation grounding score distribution across barriers.
We report Cram\'{e}r's $V$ alongside $\chi^2$ tests.
When $N$ is large, even small differences yield $p < .001$; therefore, we use $V$ to assess the magnitude of differences.

\textbf{Distribution of Content-injection Barriers.}
We first analyze the content-injection barrier distribution of citations.
To reveal the citation selection bias of GSEs, we compare the barrier distribution of top-$10$ web search results with that of GSE citations.
We use Brave Search API~\cite{brave_search_api_en} to obtain web search results. 
Since the search engines used by the target GSEs are not disclosed, we cannot replicate their exact search results.
However, Brave Search API returns results comparable to those of major search engines such as Google and Bing on standard retrieval metrics, and we therefore consider it a reasonable approximation.

We first show the barrier distribution of citations in Figure~\ref{fig:citation-by-group}, with parties grouped into ruling (rul.) and opposition (opp.).
Each cell contains two stacked bars side by side: the left bar shows the barrier distribution of web search results, and the right bar shows that of GSE citations.
We also report the mean and standard deviation of the proportion of GSE citation URLs that appear in the corresponding web search results.

The citation tendency differs across models (U.S.: $\chi^2(8) = 568.94,\, $ $p < .001,\, V = 0.328$; Japan: $\chi^2(8) = 574.80,\, p < .001,\, V = 0.250$).
GPT has the highest primary source ratio (U.S. $67.2\%$, Japan $83.1\%$) and does not cite low-barrier sources (U.S./Japan both $\leq 0.1\%$).
Gemini has the highest ratios of opponent sources (U.S. $12.4\%$, Japan $6.1\%$) and low-barrier sources (U.S. $12.9\%$, Japan $17.2\%$) among the three models.
The attack surface of Gemini is broad, with many pathways through which content injection via low-cost sources can reach the answer.
Claude has the highest mid-barrier source ratio (U.S. $59.3\%$, Japan $37.1\%$) among the three models, with primary source ratios of U.S. $16.5\%$ and Japan $53.7\%$.

The ruling and opposition distinction also affects the composition of GSE citations (U.S.: $\chi^2(4) = 1044.06,\, p < .001,\, V = 0.361$; Japan: $\chi^2(4) = 357.92,\, p < .001,\, V = 0.161$).
In the U.S., citations for ruling party questions are dominated by mid-barrier sources (Claude $78.9\%$, Gemini $58.5\%$, GPT $21.8\%$), with primary source ratios at Claude $0.9\%$, Gemini $0.9\%$, and GPT $10.3\%$.
In contrast, opposition party questions have higher primary source ratios: GPT $86.8\%$, Gemini $19.8\%$, Claude $21.3\%$.
This means that answers to ruling party questions rely more on sources with lower content-injection barriers, and the attack surface for the ruling party is broader.
In particular, for models other than GPT, we see that the primary source ratio is already low at the web search result stage.
This suggests that these models may generate search queries that are unable to retrieve primary sources.

We find that the primary source ratio in GSE citations is correlated with that in web search results.
We report in Table~\ref{tab:brave-citation-barrier-correlation-party-group}, for each model and party group, two values: Pearson correlation between the primary source share in web search results and in GSE citations ($r_P$), and the mean absolute difference between these two primary source shares (MAE) in percentage points.
We observe that the direction of the gap between party groups depends on the model: for Gemini, $r_P$ is lower for ruling parties than for opposition parties, whereas for GPT and Claude, $r_P$ is higher for ruling parties than for opposition parties.
In addition, GPT shows large MAE values, indicating that its citation composition diverges from the web search results regardless of party group, whereas Claude and Gemini exhibit much smaller MAE for U.S. ruling-party questions, meaning their citations track the web search results in that condition.

\begin{table}[t]
\centering
\footnotesize
\setlength{\tabcolsep}{1.8pt}
\caption{Correlation between primary source ratios in web search results and those in GSE citations by party group.}
\label{tab:brave-citation-barrier-correlation-party-group}
\begin{tabular}{llccc}
\toprule
\textbf{Country} & \textbf{Party group} & \textbf{GPT [$r_P$/MAE]} & \textbf{Gemini [$r_P$/MAE]} & \textbf{Claude [$r_P$/MAE]} \\
\midrule
U.S. & rul. & \textcolor{red!50!black}{$0.314$} / 22.4$\pm$28.4 & \textcolor{red!30!black}{$0.120$} / 3.5$\pm$6.7 & \textcolor{red!90!black}{$0.800$} / 0.3$\pm$1.5 \\
 & opp. & \textcolor{red!30!black}{$0.211$} / 38.6$\pm$33.4 & \textcolor{red!50!black}{$0.459$} / 14.8$\pm$12.8 & \textcolor{red!70!black}{$0.637$} / 15.8$\pm$19.0 \\
\cmidrule(lr){1-5}
JP & rul. & \textcolor{red!50!black}{$0.447$} / 49.6$\pm$21.8 & \textcolor{red!30!black}{$0.218$} / 29.1$\pm$17.7 & \textcolor{red!70!black}{$0.598$} / 27.1$\pm$17.5 \\
 & opp. & \textcolor{red!50!black}{$0.309$} / 45.0$\pm$21.9 & \textcolor{red!70!black}{$0.508$} / 19.9$\pm$15.3 & \textcolor{red!70!black}{$0.504$} / 27.0$\pm$20.6 \\
\bottomrule
\end{tabular}
\end{table}

The lower primary source ratio for ruling party questions may be caused by the larger volume of web content about ruling parties.
We additionally investigate Google search hit counts for party names: Japan's long-ruling LDP has $4.71 \times 10^{7}$ hits, far exceeding the other-party average of $8.57 \times 10^{6}$. In the U.S., the Republican Party and the Democratic Party have $1.65 \times 10^{8}$ and $3.99 \times 10^{8}$ hits, respectively.
When the total volume of web content about a party is large, the relative share of primary sources decreases, which may lower the primary source ratio in citations.
However, the U.S. Green Party has $2.19 \times 10^{9}$ hits, far exceeding the two major parties, yet its primary source ratio in GSE citations remains high. 
This counterexample suggests that publishers can suppress this relationship through their content dissemination strategies.

In contrast, user profiles do not affect the content-injection barrier distribution of citations.
We show the barrier distribution in Figure~\ref{fig:citation-by-persona} for three user profiles: Ignorant-Neutral (ign.), High-Conservative (cons.), and High-Progressive (prog.).
Each bar shows the barrier distribution of GSE citations for each user profile.
The distributions across the three profiles are nearly identical.
$\chi^2$ tests are not significant for any model in either country (U.S. overall: $\chi^2(8) = 6.62,\, p = .579,\, V = 0.020$; Japan overall: $\chi^2(8) = 10.07,\, p = .260,\, V = 0.019$; all 6 model-level tests also not significant), with all $V \leq 0.06$.
We also analyze the effect of alignment between the user profile ideology and the target party ideology: aligned (al-cons., al-prog.) and opposed (op-cons., op-prog.).
GPT shows no significant difference between aligned and opposed conditions (U.S.: $\chi^2(12) = 14.25,\, p = .285,\, V = 0.081$), selecting the same citations regardless of whether the user ideology matches the party.
Gemini (U.S.: $\chi^2(12) = 62.12,\, p < .001,\, V = 0.086$) and Claude (U.S.: $\chi^2(12) = 58.50,\, p < .001,\, V = 0.103$) show statistically significant differences, but the effect sizes are small.

\begin{table}[t]
\centering
\footnotesize
\setlength{\tabcolsep}{1.8pt}
\caption{Correlation between primary source ratio in web search results and in GSE citations by user profile.}
\label{tab:brave-citation-barrier-correlation-persona}
\begin{tabular}{llccc}
\toprule
\textbf{Country} & \textbf{User profile} & \textbf{GPT [$r_P$/MAE]} & \textbf{Gemini [$r_P$/MAE]} & \textbf{Claude [$r_P$/MAE]} \\
\midrule
U.S. & ign. & \textcolor{red!50!black}{$0.312$} / 40.1$\pm$34.8 & \textcolor{red!70!black}{$0.582$} / 11.5$\pm$11.6 & \textcolor{red!70!black}{$0.693$} / 13.3$\pm$18.7 \\
 & cons. & \textcolor{red!50!black}{$0.371$} / 36.4$\pm$33.1 & \textcolor{red!70!black}{$0.566$} / 13.3$\pm$12.7 & \textcolor{red!70!black}{$0.693$} / 12.3$\pm$18.3 \\
 & prog. & \textcolor{red!70!black}{$0.545$} / 29.5$\pm$30.6 & \textcolor{red!70!black}{$0.588$} / 12.9$\pm$13.7 & \textcolor{red!90!black}{$0.724$} / 12.5$\pm$17.6 \\
\cmidrule(lr){1-5}
JP & ign. & \textcolor{red!30!black}{$0.259$} / 46.7$\pm$22.8 & \textcolor{red!50!black}{$0.454$} / 22.0$\pm$16.2 & \textcolor{red!70!black}{$0.526$} / 27.0$\pm$20.2 \\
 & cons. & \textcolor{red!50!black}{$0.377$} / 45.1$\pm$22.1 & \textcolor{red!50!black}{$0.406$} / 22.9$\pm$16.7 & \textcolor{red!70!black}{$0.522$} / 27.5$\pm$19.8 \\
 & prog. & \textcolor{red!50!black}{$0.411$} / 46.5$\pm$20.9 & \textcolor{red!50!black}{$0.432$} / 21.6$\pm$16.4 & \textcolor{red!70!black}{$0.526$} / 26.5$\pm$19.7 \\
\bottomrule
\end{tabular}
\end{table}

We report the same correlation between the primary source ratios in web search results and GSE citations broken down by user profile in Table~\ref{tab:brave-citation-barrier-correlation-persona}.
Within every (country, model) cell, $r_P$ and MAE are almost the same across the three user profiles.
The per-model correlations hold across all three user profiles as well.
This suggests that user profiles affect neither the citation barrier distribution itself nor the size of its difference from the web search results.

\begin{figure}[t]
  \centering
  \includegraphics[width=\columnwidth]{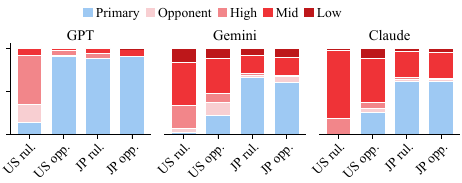}
  \caption{Domination score distribution by party group. Each bar shows the barrier-wise domination score.}
  \label{fig:wc-by-group}
\end{figure}

\begin{figure*}[t]
  \centering
  \includegraphics[width=\textwidth]{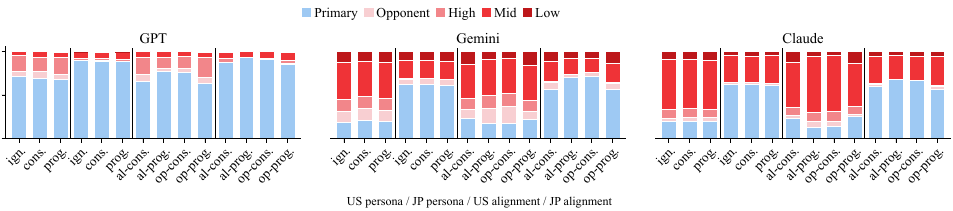}
  \caption{Domination score distribution by user profile. Each bar shows the barrier-wise domination score for each user profile.}
  \label{fig:wc-by-persona}
\end{figure*}

\textbf{Distribution of Domination Score.}
We next analyze the domination score distribution across barriers.
We show the distribution by model and party group in Figure~\ref{fig:wc-by-group}.
As with barrier distribution, model choice and the ruling/opposition position again govern the distribution (model: U.S. $\chi^2(8) = 34120,\, V = 0.356$, Japan $\chi^2(8) = 54892,\, V = 0.229$; party group: U.S. $\chi^2(4) = 59662,\, V = 0.386$, Japan $\chi^2(4) = 31743,\, V = 0.144$; all $p < .001$).

Across all models and both countries (ruling/opposition combined), the primary source domination ratio exceeds the citation ratio.
GPT shows ${+}3.9$pt in the U.S. and ${+}7.0$pt in Japan; in Japan, Gemini shows ${+}13.1$pt and Claude ${+}8.5$pt.
This means that the text span of the answer attributed to each primary source citation is longer than that of citations from other barriers.
However, the ruling/opposition gap depends strongly on model and country.
In the U.S., the primary domination is higher for opposition questions (GPT ${+}76.6$pt, Claude ${+}25.0$pt, Gemini ${+}20.5$pt), and the U.S. ruling-party primary domination for GPT drops to $14.6\%$; in Japan, the gap stays within $6$pt for all models.
For low-barrier sources, Gemini's domination in Japan substantially exceeds the other models (Gemini $9.8\%$ vs.\ Claude $4.3\%$, GPT $0.1\%$).

We show the domination score distribution by user profile in Figure~\ref{fig:wc-by-persona}.
User profile differences are negligible (U.S. $\chi^2(8) = 301,\, V = 0.019$, Japan $\chi^2(8) = 377,\, V = 0.011$; all $p < .001$).
Together with barrier distribution (all $V \leq 0.06$), user profiles do not affect the attack surface in either metric.

\begin{figure}[t]
  \centering
  \includegraphics[width=\columnwidth]{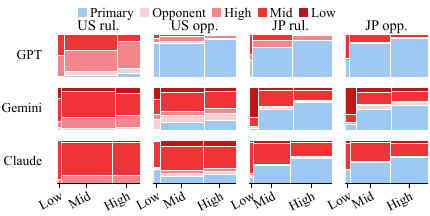}
  \caption{Grounding distribution by party group. In each cell, the width shows the count share of each grounding bin, and the height shows the barrier distribution within that bin.}
  \label{fig:grounding-by-group}
\end{figure}

\begin{figure*}
  \centering
  \includegraphics[width=\textwidth]{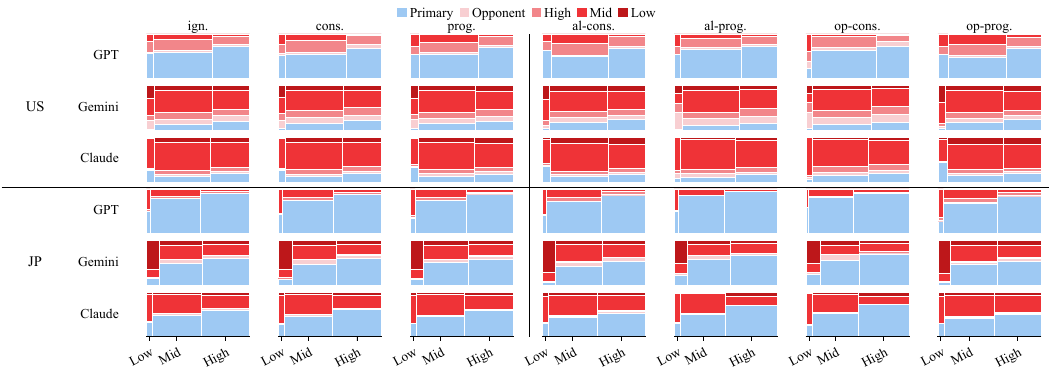}
  \caption{Grounding distribution by user profile. Each cell shows the grounding bin barrier distribution for each user profile.}
  \label{fig:grounding-by-persona}
\end{figure*}

\textbf{Distribution of Citation Grounding.}
We next analyze the citation grounding score distribution across barriers.
We discretize grounding scores with thresholds $(\tau_L, \tau_H)$ into Low ($<\tau_L$), Mid ($[\tau_L, \tau_H)$), and High ($\geq \tau_H$).
Because the distribution of SBERT cosine similarity can differ by model and language~\cite{shibayama2021japanesesbert}, we calibrate the thresholds separately for English (U.S.) and Japanese (Japan).
For calibration, we use the semantic textual similarity (STS) benchmarks STS-B~\cite{cer2017stsb} for English and JSTS~\cite{kurihara2022jglue} for Japan.
In these datasets, each sentence pair is assigned a human similarity score from $0$ (unrelated) to $5$ (identical in meaning).
SBERT maps sentences into a vector space and compares their meaning using cosine similarity.
Its sentence embeddings are commonly evaluated by their correlation with human scores on STS benchmarks~\cite{reimers2019sbert}.
The empirical relationship between human scores and cosine similarity on STS data allows us to set language-specific thresholds.

We define $\tau_H$ as the median cosine similarity of pairs with human score $\geq 4$, and $\tau_L$ as the median cosine similarity of pairs with human score $\geq 3$.
Under this calibration, $\tau_H$ serves as a conservative threshold for near-paraphrase similarity, whereas $\tau_L$ serves as a threshold for substantial shared content.
As a result, we obtain $(\tau_L^{\text{U.S.}}, \tau_H^{\text{U.S.}}) = (0.80, 0.90)$ and $(\tau_L^{\text{Japan}}, \tau_H^{\text{Japan}}) = (0.70, 0.85)$.
In SBERT-based sentence similarity studies, values such as $0.8$ and $0.9$ are commonly used to detect paraphrases and highly similar sentences~\cite{kabir2024banglaembed,tumre2025improved}, which agrees with our English thresholds.
On the other hand, Japanese SBERT may give lower cosine similarity than English models even for synonymous or highly similar pairs~\cite{shibayama2021japanesesbert}, so we use lower thresholds for Japan.

High means the answer is strongly grounded in the source, Mid means the core content is partly shared, and Low means only a weak semantic link.
For Japanese questions, $86\%$ of citation sources are in Japanese and $14\%$ in English, while for English questions $100\%$ are in English.
Since a single language is dominant within each country in questions and citations, language-specific calibration is reasonable for consistent comparison.

We show the barrier distribution for each grounding bin in Figure~\ref{fig:grounding-by-group}.
Each cell is a Marimekko chart: the width along the x-axis shows the count share of each grounding bin (Low/Mid/High), and the y-axis shows the barrier distribution within each bin.
Except for U.S. Claude, the primary source ratio increases as the grounding bin goes from Low to Mid to High.
For example, in Japan, GPT shows primary ratios of $42.1\%$ (Low bin), $75.6\%$ (Mid), and $89.2\%$ (High).
The magnitude of this increase differs by country and model: all three models in Japan show steep increases (Low$\to$High by ${+}30$ to ${+}47$pt), while in the U.S. GPT and Gemini show only moderate increases (${+}15$ to ${+}17$pt), and Claude is non-monotonic with Low$=27.5\%$, Mid$=12.9\%$, High$=18.2\%$.
Except for this case, citations with high grounding scores are concentrated in primary sources, and citations with low grounding scores are concentrated in low-barrier sources.
That is, primary sources with high barriers are faithfully reflected in answers, while sources with lower barriers contribute less to the answer text and are incorporated in a form that diverges more from the original content.

This tendency is not uniform across ruling and opposition parties.
Answers to ruling party questions have lower High-bin ratios.
In the U.S., the High-bin ratio for ruling party questions is GPT $28\%$, Gemini $30\%$, Claude $33\%$, while for opposition party questions it is GPT $38\%$, Gemini and Claude $39\%$ ($\chi^2(2) = 1091,\, p_H < .001,\, V = 0.094$ with Holm correction).
Given that ruling party questions also have lower primary source ratios in the barrier distribution, their answers rely more on citations in the Mid and Low grounding bins.

The effect size of user profile on the grounding distribution is also negligible (U.S. $V = 0.012$, Japan $V = 0.009$).
GPT (U.S.) and Claude (U.S. and Japan) are not significant ($p_H \geq .05$), while the other combinations show statistically significant differences ($p_H < .001$); however, all have $V \leq 0.022$, meaning the effect is negligible.
This suggests that the potential attack surface of GSEs does not depend on who the user is, but on which GSE model is used and which party the question is about.

\subsection{Summary}
While LLM outputs can vary with the user's political ideology, we observe no consistent effect on GSE citation behavior across the three models and both countries.
What shapes the differences in citation behavior is not \emph{who is asking the question} but \emph{which model answers about which party}.
Model choice produces large differences in the barrier distribution, ruling-party questions have lower primary source ratios than opposition-party questions, and in the U.S. the domination and grounding distributions skew in the same direction.
That is, the attack surface of GSEs is characterized by the model and the target party, not by the user's ideology.
\section{Discussion}
\label{sec:discussion}
We discuss the ideal citation balance between primary and secondary sources in GSEs.
For closed-ended questions in the political domain, increasing citation proportions of primary sources is desirable.
However, prior studies claim that primary providers may present only excessively positive aspects in content shown to users~\cite{kluver2016setting,leung2015impression}.
When social doubts arise about the reliability of primary sources, secondary sources are appropriate to include in citations.
For instance, regarding product reviews, not only official product websites but also secondary sources such as reviews by others should be included in citations of answers.
Some domains and tasks require secondary sources alongside primary sources, and it is not always optimal to maximize citation proportions of primary sources.
Accordingly, it is desirable to define an ideal balance between primary and secondary sources for each task and domain, and to evaluate the citation tendencies of GSEs based on it.

We discuss two limitations in our study.
First, we target closed-ended questions in the political domain in the U.S. and Japan rather than conducting a large-scale cross-domain study.
We also limit profile dimensions to political knowledge and ideology, excluding attributes such as age, gender, and education to contain experimental cost.
Future work will extend to more models, questions, user attributes, and domains such as health and finance.

Second, our category classification approach has limited granularity in distinguishing between web source types within each publisher attribute category.
Content-injection barriers depend on our publisher attribute classification.
However, our method does not account for cost differences, such as publisher selection processes or peer review processes on individual pages.
Content-injection barriers vary between peer-reviewed journal papers and preprint papers, and between different newspaper publishers.
By considering these differences within the attribute groups, future work can capture content-injection barriers at a finer resolution.

\section{Conclusion}
\label{sec:conclusion}
We proposed a framework to characterize the potential attack surface of GSEs against poisoning attacks in the political domain.
To build our framework, we introduced the content-injection barrier, a novel evaluation criterion.
Our experiments on three major GSEs in the U.S. and Japan showed that (a) each model has a distinct attack surface; (b) the web search functionality of GSEs (e.g., query generation by LLMs) shapes the attack surface; (c) ruling parties have a broader attack surface than opposition parties; and (d) user profiles do not affect the potential attack surface.
Our findings suggest that the potential attack surface depends on the model and the target party, not on the user.

\section*{Acknowledgments}
We thank Prof. Koichi Moriyama at Keio University for his helpful feedback.
This study was supported by a donation from Yahoo Japan Corporation to the Ataka Laboratory at Keio University Shonan Fujisawa Campus.

\section*{GenAI Usage Disclosure}
We used Claude Opus 4.7 to assist with visualization scripts, code refactoring, and language refinement of author-written drafts.
All outputs were reviewed and revised by the authors, who are responsible for the final content.

\bibliographystyle{acm}
\bibliography{./bib/cites}


\end{document}